**Metagenomic analysis revealed significant changes in cattle rectum microbiome and antimicrobial resistome under fescue toxicosis**

Yihang Zhou[1†]

[1]School of Life Sciences and Technology, Tongji University, Shanghai, 201613, China

[†]co-corresponding author:

Yihang Zhou
E-mail: youngjoe@tongji.edu.cn

**Abstract**

Fescue toxicity causes reduced growth and reproductive issues in cattle grazing endophyte-infected tall fescue. To characterize the gut microbiota and its response to fescue toxicosis, we collected fecal samples before and after a 30-days toxic fescue seeds supplementation from eight Angus×Simmental pregnant cows and heifers. We sequenced the 16 metagenomes using the whole-genome shotgun approach and generated 157 Gbp of metagenomic sequences. Through *de novo* assembly and annotation, we obtained a 13.1 Gbp reference contig assembly and identified 22 million microbial genes for cattle rectum microbiota. We discovered a significant reduction of microbial diversity after toxic seed treatment ($P<0.01$), suggesting dysbiosis of the microbiome. Six bacterial families and 31 species are significantly increased in the fecal microbiota ($P$-adj$<0.05$), including members of the top abundant rumen core taxa. This global elevation of rumen microbes in the rectum microbiota suggests a potential impairment of rumen microbiota under fescue toxicosis. Among these, *Ruminococcaceae bacterium P7,* an important species accounting for ~2% of rumen microbiota, was the most impacted with a 16-fold increase from 0.17% to 2.8% in feces ($P<0.01$). We hypothesized that rumen *Ruminococcaceae bacterium P7* re-adapted to the large intestine environment under toxic fescue stress, causing this dramatic increase in abundance. Functional enrichment analysis revealed that the overrepresented pathways shifted from energy metabolism to antimicrobial resistance and DNA replication. In conclusion, we discovered dramatic microbiota alterations in composition, abundance, and functional capacities under fescue toxicosis, and our results suggest *Ruminococcaceae bacterium P7* as a potential biomarker for fescue toxicosis management.

**Background**

Tall fescue, *Lolium arundinaceum (Schreb.) Darbysh,* is a widely adapted, cool-season perennial grass grown on over 14 million hectares in the Southeastern US alone [1], feeding over 12 million beef cattle [2]. Most tall fescue plants are infected with an endophyte, *Epichloë coenophiala* [3], which lives in a mutualistic symbiosis with the plant. This mutualism produces secondary metabolites that protect the plant against abiotic and biotic stress [4-8], and increases stand longevity [9]. However, this symbiotic relationship also results in high concentrations of ergot alkaloids (i.e., ergovaline), causing reduced growth and reproductive problems in cattle grazing endophyte-infected tall fescue, which are known as fescue toxicity [10-12]. The economic losses due to fescue toxicity to the forage-based livestock industry are estimated to be close to 3.5 billion US dollars each year [13].

The negative effect of tall fescue feeding on cattle growth is affected by the host genotype. Genomic approaches have great potential to identify and select resistant animals. Several studies have investigated genetic markers with a significant contribution to fescue toxicity tolerance. Among these, a polymorphism within the XK-related 4 gene was discovered to associate with serum prolactin levels [14], and an intronic SNP (single nucleotide polymorphism) was identified in the dopamine receptor D2 gene [15] in beef cows grazing endophyte-infected tall fescue.

The host metabolism was significantly affected under fescue toxicosis. Previous studies detected a significant tryptophan and lipid metabolism disruption [16] and dramatic shifts in amino acid metabolism in the plasma [17], among the main consequences of ergovaline consumption. In ruminants, the digestive function is dependent on the fermentation mostly by the rumen microbiota, and the gut microbes to a lesser extent [18, 19]. The gut microbiota is directly related to important biological functions such as nutrition and digestive health [20, 21], affecting almost every aspect of animal physiology, including metabolism, immunity, inflammation, and behavior [22]. The rumen microbiome has been studied extensively [23-32], and it is shown to be associated with feed efficiency in beef cattle [33-37]. The impact of fescue toxicity on the rumen microbiome was previously reported, summarizing the metabolic and physiologic impact on cattle and discussing different mitigation strategies [38]. However, the fecal/rectum microbiota and its response to ergovaline are not well understood, with only 16S rDNA gene sequencing studies [17, 39, 40].

To fill in this gap in knowledge, we performed metagenomic analyses in cattle rectum microbiota before and after the consumption of endophyte-infected tall fescue seeds (before treatment and after treatment in short). We selected the rectum microbiota specimen because it is more practical to obtain fecal samples from the field compared to rumen samples, which is an advantage in developing biomarkers to assessing the tall fescue toxicosis. Previous studies using 16S metagenomic sequencing in fecal microbiota [17, 39, 40] have identified a significant effect of fescue treatment on the fecal microbiota diversity, but the resolution is mostly at the microbial genus level. Animal gut microbiota is extremely complex, with more than 1,000 species of microbes. Beyond the taxonomy composition, extensive horizontal gene transfer (HGT) [41] and copy number variation at the species and strain level add additional layers of complexity [42]. Thus, the 300 bp amplicons from the 16S rDNA gene sequencing approach lack finer resolution into the functional genomic complexity inherent in microbiomes. To detect microbial species, strain, pathway, and gene-level changes in the gut microbiota in response to fescue toxicity, we applied whole-genome shotgun (WGS) metagenomic sequencing in fecal samples, with the ability to identify bacteria composition at the species level, infer individual bacterial strain abundance alterations based on within-species variations, and detect enrichment of gene networks and metabolic pathways [43].

## Methods

### Animal selection and maintenance

Eight Angus × Simmental pregnant cows and heifers from were selected for tall fescue toxicosis analysis. Based on the fescue toxicity tolerance index (Table S1), animals were assigned to two groups: susceptible (n = 4) and tolerant (n = 4). The toxicity tolerance indices were determined based on a genetic test for tolerance to fescue toxicity [44-46], which provides a tolerance index ranging from zero star (most susceptible) to five stars (most tolerant).

### Endophyte-infected tall fescue seed feed supplement treatments.

The tolerant and susceptible animal groups have uniform average body weight (BW; 562 ± 14 kg; $P$ = 0.69) within groups. To ensure a daily dose of ergovaline that produces specific signs of fescue toxicity in beef cattle, we formulated the diet to achieve an individual daily intake of 20 µg of ergovaline/kg body weight/day. Specifically, each animal of both groups received treatment twice per day, and each treatment includes all animals received the following diet: 2.55 kg of endophyte-infected K31 tall fescue seeds, 1.16 kg of soybean hulls, 1.16 kg of corn gluten feed, and 0.24 kg molasses in a dry matter basis in addition to *ad libitum* Bermudagrass hay. The ergovaline content in fescue seeds was 5,000 ppb on a DM basis, as measured by high-performance liquid chromatography (HPLC; detection limit = 50 ppb and CV = 7%) [47]. Animals had free access to water and mineral supplementation blocks during the adaptation period and the trial. The study started when the animals were on average six months pregnant and received K31 fescue seeds for 30 days (Figure 1A). Individual body weight was obtained on a weekly basis (Table S2).

**Fecal sample collection, microbial DNA extraction, and rectum metagenomic sequencing**

Fecal samples were obtained at day 0 (7/1/2019) and day 29 (7/29/2019) of the trial at 0700 using single-use fecal loops and immediately placed on liquid nitrogen for posterior storage in -80°C freezer until laboratory analysis. Fecal DNA samples were extracted using the Qiagen Allprep PowerFecal DNA/RNA kit (QIAGEN, MD). The homogenization step was conducted using the Qiagen PowerLyzer24 instrument (QIAGEN, MD). The DNA concentrations were measured by Qubit fluorometer (Invitrogen) with dsDNA high-sensitive assay kit. 2 µg of genome DNA was fragmented by M220 Focused-ultrasonicator with a 500bp targeted insert size. Metagenomic sequencing libraries were made using the NEBNext Ultra II DNA Library Prep Kit for Illumina, following manufacture's protocols. The size range of the final library insert size was 500-600 bp.

**Quality control and preprocessing of metagenomic reads**

On average, we generated 10.1 billion base-pairs (bp) of sequences per sample for each of the 16 metagenomes (Table S3). The metagenomic reads were quality checked using FastQC (v0.11.9) [48]. The paired-end reads were trimmed to remove Illumina adapter sequences and low-quality bases by Trimmomatic (v0.36) [49]. High-quality filtered reads were mapped to the cattle reference genome (GenBank: GCA_002263795.2) using Burrows-Wheeler Aligner (BWA) (v0.7.17-r1188) [50], to remove any host contaminations (Table S3). The survived reads were aligned to the viral genome database downloaded from National Center for Biotechnology Information (NCBI) to remove the viral sequences (Table S3) [51]. The bacterial reads were extracted using SAMtools (v1.6) [52, 53] and BEDTools (v2.25.0) [54] for subsequent analysis.

**Cattle rectum metagenome assembly, taxonomy assignment, and microbial gene annotation**

In order to obtain high-quality *de novo* assembly, after the quality control pipeline, we joined the paired-end (PE) reads using PEAR (v0.9.11) [55] to obtain longer merged reads. We assembled the merge single-end reads and PE reads into metagenomic contigs using MEGAHIT (v1.1.2) [56] based on the de Bruijn graph [57] approach. Redundant contigs were removed through clustering using CD-HIT (v4.7) [58] with a 97% sequence identity threshold. Taxonomy assignments for these non-redundant metagenomic contigs were performed using Kaiju (v1.7.3) [59] at species, genus, family, order, class, and phylum levels. Microbial genes were predicted from assembled non-redundant contigs using meta-gene predictor MetaGeneMark (v3.38) [60, 61].

**Cattle rectum microbiome diversity analysis**

The alpha-diversity was analyzed using R package vegan v2.5.7 [62] with Shannon index [63] (Figure 1B). The beta-diversity was analyzed based on the Bray-Curtis dissimilarity [64] and Jaccard index in R. The Principal Coordinates Analysis (PCoA) was performed in R [65] (Figure 1C).

**Metagenomic analysis of taxonomy abundance changes in response to tall fescue toxicosis**

The high-quality filtered paired-end reads from each metagenome were aligned to the rectum metagenomic contigs we assembled to generate mapping counts tables (Table S4). With the counts tables and contig taxonomy annotation, we computed the relative taxonomy frequency using custom Python (v3.7.7) scripts (Data S1-S4; Table S5 and S6). The relative frequency bar-plots were generated using the R package ggplot2 v3.3.3 [66] and reshape2 v4.0.5 [67] (Figure 2). To assess the statistical significance of the differential abundance, two-sided Wilcoxon signed-rank tests were performed in R using package ggpubr (Table S7 and S8). The heatmap plots were generated using R packages pheatmap (v1.0.12) and grid (v4.0.5). The *q*-values were calculated using the qvalue (v2.22.0) package in R.

**Identification of the most discriminative features before vs. after tall fescue treatment.**
We performed linear discriminant analysis Effect Size (LEfSe v1.1.1) analysis to determine the most relevant features that explain the differences between groups [68] (Figure 3). Relative taxonomy composition frequencies were used to determine the characteristic microbial taxa in the LEfSe pipeline. HUMAnN2, a pipeline for profiling the abundance of microbial pathways [69], was used to generate the functional pathway profiles in the 16 metagenomes. The most featured functional pathways were identified based on HUMAnN2 output using LEfSe. The alluvial plot was generated using the R package ggalluvial v0.12.3 [70] (Figure 3).

**Relative abundance analysis of Ruminococcaceae bacterium P7 in the rectum microbiota**
We used bar plots and heatmap to illustrate the change of *Ruminococcaceae bacterium P7* relative frequency before and after treatment (Figure 4). Phylogenetic analysis of microbiome abundance was performed hierarchically from species to phylum level based on the counts tables. We used Krona (v2.8) [71] to visualize the phylogenetic relationship and abundance proportions of *Ruminococcaceae bacterium P7* within the Firmicutes phylum and the Ruminococcaceae family (Figure 5). To examine the changes of other Ruminococcaceae members, we compared the relative abundance of 17 *Ruminococcaceae bacterium* species before and after treatment using a heatmap (Figure 4).

**Comparative metagenomic analysis of rectum microbiome with rumen microbiome**

To compare our rectum metagenomic data with the cattle rumen microbiota, we used a high-quality rumen metagenome reference [72], which was assembled using 4,941 rumen metagenomes (European Nucleotide Archive accession number PRJEB31266). The rumen reference genome, as well as the top 41 core rumen microbiota genera and top 31 abundant rumen microbiota families, are adopted from Stewart *et al.* [72]. To measure the similarity between the cattle rectum and rumen microbiota, we aligned our PE reads to the rumen reference. We used Python packages matplotlib (v3.3.4) [73] and matplotlib_venn (v0.11.6) to draw the Venn plot (Figure 6). The relative abundance of the microbial taxa was determined by counts per million mapped reads (CPM), which was calculated using custom Python scripts [74].

**qPCR validation of microbial composition changes**

To confirm the *Ruminococcaceae bacterium P7* abundance changes detected in the metagenomic data, we designed primers for quantitative PCR experiments to target two single-copy genes, *rnhB*, (forward: 5'-TTTGCCGGATCTTGATATTGATGC-3'; reverse: 5'-TTTTCAAAACCATACTGCGGAA-3'), and *recR* (forward: 5'-GGCGCACACCCGTATCCAC-3'; reverse: 5'- ACACCCTGAAACTCGTGCGTA-3'). qPCR experiments were performed with SYBR Green (Thermo Fisher, Cat No. A25776) in 96-well plates on a Bio-Rad C1000 Touch Thermal Cycler with CFX96 Real-Time PCR Detection Systems. Initial analysis was done using the machine software. Metagenomic contig cattle_contig_000004026845 (annotated as *Butyrivibrio sp. AE2032*) was selected as the control because stable relative abundance was observed across all 16 metagenomes. qPCR experiments were performed for this control species in all 16 fecal DNA samples. Relative quantification was performed based on ratios of *Ruminococcaceae bacterium P7* and control species abundance. Two technical replicates were included for each sample.

**Enrichment of functional categories and pathways before and after tall fescue treatment.**

The HUMAnN2 pipeline was used to profile the function and pathways of all samples [69]. The pathways were also annotated using the KEGG database [75]. LEfSe for features identifications were performed based on the KEGG Ortholog tables and the relative abundance quantified by normalized CPM. The KEGG Orthologs were grouped into KEGG BRITE hierarchies to investigate high-level functional terms.

**Analysis of antimicrobial resistome before and after tall fescue treatment.**

We used ABRicate (v0.8.13) to mass screen our rectum microbial contigs for antimicrobial resistance genes (ARGs) using the Resfinder database [76]. The ARGs were further grouped into ARG families using the Comprehensive Antibiotic Resistance Database (CARD) [77]. We extracted the reads aligned to the annotated ARGs to quantify the relative abundance of each ARG family with custom Python scripts (Figure 7 and Data S5).

## Results

**A comprehensive assembly of cattle rectum microbiome using WGS metagenomic data.**

In this study, we generated a total of 157 Gbp of metagenomic sequences. On average, 65.4 million 150-bp reads were generated per sample for each of the 16 cattle rectum microbiota using WGS-metagenomic sequencing (Figure 1A and Table S3; see Methods). 2.20% of adapter and low-quality bases, 6.31% of host sequences, and 0.03% of viral reads, were excluded from subsequent analysis (Table S3). After filtering, a total of 104.4 Gbp of high-quality microbial sequences were assembled *de novo* into cattle rectum microbial contigs. The non-redundant reference assembly consists of 16,580,560 contigs, ranging from 400 bp to 406,369 bp with an N50 of 787 bp. The total length of the rectum reference contigs is 13.1 Gbp. The metagenomic reads from each metagenome were aligned to our cattle rectum microbiota reference, and the average mapping rate is 91.7% (Table S4). Taxonomic assignments and gene annotation were performed on the reference contigs, and 21,950,894 non-redundant microbial genes were identified (total length = 9.8 Gbp; N50 = 540 bp).

**Significant reduction in cattle rectum microbiota diversity in response to endophyte-infected tall fescue seed supplement.**

After the 30-day consumption of endophyte-infected tall fescue seed, the alpha diversities in the rectum microbiota were significantly decreased for both species-level abundances (Figure 1B, $P = 0.008$, Wilcoxon signed-rank test) and genus-level abundances (Figure 1C, $P = 0.016$). Our results indicate a strong reduction in rectum microbiome complexity after tall fescue seed treatment. Beta diversities were analyzed and plotted for the rectum metagenomes before and after treatment. Principal coordinates analysis (PCoA) identified significant separation for the two groups under both the Bray-Curtis dissimilarity measurement (Figure 1D, $P = 0.002$, PERMANOVA) and Jaccard distance (Figure 1E, $P = 0.004$, PERMANOVA). Collectively, the microbial community diversity analyses revealed different compositions between metagenomes before and after tall fescue treatment (Figure 1).

**Significant alterations in phylum abundance: Firmicutes increased in the cattle rectum microbiota under tall fescue toxicosis.**

Among the rectum reference microbial contigs we assembled, 70.0% were assigned to a specific phylum, and 50.3% have species-level annotations. The most abundant phyla in cattle rectum microbiota are Firmicutes and Bacteroidetes (Figure 2A), which account for ~80% of the total abundance, which is consistent with previous studies using 16S rRNA gene data [40]. After examining the overall phylum-level composition of before and after treatment metagenomes for the eight animals, we discovered a 7.6% enrichment of Firmicutes ($P = 0.002$), an 18.1% reduction of Proteobacteria ($P = 0.008$), and a 15.6% decrease of Actinobacteria ($P = 0.04$) after consuming toxic tall fescue seeds (Figure 2A-B). There was a trend of more reduction in Firmicutes for susceptible animals (10.1% reduction on average) than individuals with a tolerant T-snip genotype (5.1% reduction on average), but the results did not achieve statistical significance ($P = 0.31$. Mann-Whitney U-test, Figure S1). To further investigate the difference abundance before vs. after treatment, we profiled the relative frequencies at bacterial family (Figure 2C) and species levels (Figure 2D). Ruminococcaceae, Lachnospiraceae, Bacteroidaceae, Clostridiaceae, and Rikenellaceae are the top 5 most abundance families (Figure 2C), which together account for 53.8% of the total abundance. The top 5 most abundant species are unclassified *Clostridiales bacterium* (NCBI taxonomy ID 2044939), *Firmicutes bacterium CAG*

*11*, unclassified *Ruminococcaceae bacterium* (NCBI taxonomy ID 1898205), *Sarcina sp. DSM 11001*, and *Ruminococcaceae bacterium P7* (Figure 2D). *Clostridiales bacterium* and *Ruminococcaceae bacterium* are defined as species in the NCBI annotation database, but they are a collection of unassigned bacteria within the higher taxonomy unit rather than a distinct species. For the sake of bioinformatic analysis using NCBI annotation, they were treated as individual species. Notably, a significant increase of *Ruminococcaceae bacterium P7* was eminent in the heatmap (Figure 2D).

**Top 10 most featured families/species that discriminate the cattle rectum metagenomic profiles before and after toxic tall fescue treatment.**

Linear discriminant analyses were performed for the metagenomic profiles before and after treatment, and we identified the top 10 most featured families (Figure 3A) and species (Figure 3B). Among these, the overrepresented bacterial families after treatment are Ruminococcaceae (LDA score=4.67), Lactobacillaceae (3.62), Peptostreptococcaceae (3.55), Selenomonadaceae (LDA score=3.37), and Aerococcaceae (LDA score=3.10, Figure 3A). Lentisphaerae (LDA score=3.81), Desulfovibrionaceae (3.56), Rikenellaceae (3.53), Eggerthellaceae (3.48), and Victivallaceae (3.30) were significantly decreased after treatment ($P < 0.05$, Wilcoxon signed-rank test). At the species level, the top five most abundant species after treatment based on LDA scores are *Ruminococcaceae bacterium P7* (16-fold increase), *Sarcina sp. DSM 11001* (48.5% increase), *Ruminococcus bromii* (3-fold increase), *Lactobacillus ruminis* (14-fold increase), and *Ruminococcus sp. JE7A12 P7* (6-fold increase). Among the top 5 decreased species annotation, only two are individual species (*Clostridium sp. CAG 448* and *Anaerotruncus sp. CAG 390*). The rest three are unclassified Firmicutes/Ruminococcaceae/Clostridiales (Figure 3B). The alluvial plot revealed that the top 5 decreased species have a much lower degree of change (15.2% overall all abundance reduction) compared to the top 5 increased species (215.8% overall all abundance increase, Figure 3C). The same pattern holds when all significantly increased (Figure 3D) and decreased species (Figure 3E) are examined.

**The dramatic increase of a Firmicute species, *Ruminococcaceae bacterium P7*, is a hallmark of cattle rectum microbiota under tall fescue toxicosis.**

As the most featured species discriminating before and after treatment rectum microbiome, *Ruminococcaceae bacterium P7* really stands out when species-level relative frequencies are investigated (Figure 4A), with a ~16-fold enrichment after treatment ($P = 0.0078$, Figure 4B). This significant abundance change is consistent across all eight individuals (Figure 4A). Because *Ruminococcaceae bacterium P7* lacks genus annotation (unclassified Ruminococcaceae, Taxonomy ID 1200751), we checked the abundance profiles of related species in the same family. Among the 17 *Ruminococcaceae bacterium* species, *Ruminococcaceae bacterium P7* is the only one that displayed a significant increase after tall fescue seed consumption (Figure 4C), indicating an interesting species-specific effect.

Over 50% of microbes in cattle rectum microbiota are Firmicutes (Figure 2A), and Ruminococcaceae is the most abundant family (Figure 2C and 4A-B), suggesting its functional importance in the gut. Strikingly, Ruminococcaceae is also the most changed family between the before and after groups (Figure 3A). Through phylogenetic analysis of microbial composition in Firmicutes, we discovered the change of Ruminococcaceae is largely driven by ~16-fold increase of the bacterial species *Ruminococcaceae bacterium P7* (Figure 5A-B). Among the top 10 most feature species (Figure 3B), *Sarcina sp. DSM 11001* increased from 4% to 6% abundance in Firmicutes (Figure 5A-B), which also contributed to the overall enrichment (Figure 3A). The two most decreased species *Clostridium sp. CAG 448*, and *Anaerotruncus sp. CAG 390* displayed 30.2% and 27.6% abundance reduction in Firmicutes, respectively (Figure 5A-B). Zooming in to Ruminococcaceae, many other bacterial members in the family (i.e., *Ruminococcus sp. CAG:177*, *Ruminococcus flavefaciens*, and *Faecalibacterium prausnitzii*) did not show significant abundance change (Figure 5C-D).

**Ruminococcaceae is a core rumen microbiome family, and a global increase of rumen core taxa was observed in the rectum microbiome after toxic tall fescue treatment.**

To investigate the bacterial composition and abundance differences between rectum and rumen microbiota, we aligned our metagenomic reads to the rumen reference metagenomes from Stewart *et al.* [72], and discovered that 34.0% of the rectum microbial reads could be mapped to the rumen reference (Figure 6A), suggesting more than a third of the rumen microbes could be found in the rectum microbiota. After the consumption of toxic fescue seeds, we observed a 10.1% increased of rumen-mapped reads, which is marginally significant with a *P*-value of 0.078 (Figure 6A). Based on this result, we conclude that there is a trend of increased rumen microbial abundance in the rectum microbiota after treatment. To further examine this observation in detail, we quantified the abundance of 41 core rumen genera [72] in the rectum data, and identified a significant bias toward upregulation of rumen core genera after treatment (Figure 6B). 65.9% (27/41) of the core genera increased after treatment, and the magnitude is significantly higher (up to over 300%) compared to the downregulated genera (<50%). After ranking the rectum microbiota genera, we found that 16 of the top 41 most abundant rectum microbiota genus are shared in the rumen microbiota (Figure 6C). When we studied the diversity based on rumen-mapped reads only, the alpha-diversity decreased after treatment ($P = 0.04$, Wilcoxon signed-rank test, Figure 6D), the beta-diversity also demonstrated significant dissimilarity between the before- and after-group ($P = 0.02$, PERMANOV, Figure 6E), which is consistent with the results using all rectum reads (Figure 1). Because *Ruminococcaceae bacterium P7* lacks genus annotation, we compared the abundance profile of the top 31 rumen microbial families from Stewart *et al.* [72], with their corresponding rectum abundance before and after treatment. In general, highly abundant rumen families also have a higher rank in the rectum profile (Figure 6F), with 6 of the top 8 core families shared between the two, indicating their importance in rumen function. Among them, Ruminococcaceae is the only one within the list of top 10 families which discriminate the before and after treatment groups (Figure 3A).

**Functional category enrichment analysis revealed increased antimicrobial resistance (AMR) capacity after toxic tall fescue treatment.**

Linear discriminant analyses revealed the top 10 most featured KEGG terms between the before- and after-treatment group. Interestingly, energy metabolism-related orthologs are overrepresented before treatment, whereas antimicrobial resistance genes (ARGs) related orthologs are enriched in the after-treatment metagenomes (Figure 7A). We summarized all annotated KEGG orthologs according to the BRITE functional hierarchies, and identified 17 out of 28 KEGG BRITE terms increased significantly after treatment ($P < 0.05$, Wilcoxon signed-rank test, Figure 7B). Eight KEGG BRITE terms have at least a 2-fold increase in abundance, including transporters, ARGs, chromosome and associated proteins, DNA replication proteins, lipid biosynthesis proteins, glycosyltransferases, peptidoglycan biosynthesis and degradation proteins, and bacterial motility proteins (Figure 7B). To determine which ARGs increased in response to fescue toxicity, we investigated the antimicrobial resistome in the before-treatment and after-treatment metagenomes through ARG prediction and relative abundance profiling (see Methods). 89 ARGs were identified in the before-treatment microbial contig assembly, and 94 were discovered after treatment, suggesting an increase in the number of ARGs under fescue toxicosis. We binned the ARGs into 21 categories and quantified their normalized abundance in each individual metagenome (Figure 7C). Three gene families, tetracycline-resistant ribosomal protection proteins, major facilitator superfamily (MFS) antibiotic efflux pumps, and ABC-F ATP-binding cassette ribosomal protection proteins, are significantly increased after treatment ($P < 0.05$, Wilcoxon signed-rank test, Figure 7D).

**Discussion**

To improve the feed efficiency and reduce greenhouse gas emissions, cattle rumen microbiota were comprehensively studied at WGS metagenomic level [31, 32], metatranscriptomic level [28, 29], as well as individual genome sequencing of rumen microbes [30, 78]. In contrast, there is a paucity of fecal/rectum microbiome studies using WGS metagenomic approach. In our study, we sequenced and assembled rectum metagenome from 16 individual samples, and identified 22 million non-redundant microbial genes. The cattle rectum microbiota is more complex than humans [79], rodents [80, 81], canine [82], or swine [83]. Furthermore, 34% of the rectum microbial sequences could be mapped to the rumen metagenomes (Figure 6A). A previous study, in which 16S rDNA amplicon sequencing was performed, found that fecal/rectum OTUs in Nelore cattle are highly similar to the cecum and colon profiling, which could reflect the microbial composition in the large intestine [78]. The most highly represented phyla in our data are Firmicutes (54.4%) and Bacteroidetes (29.6%), and the most abundant families are Ruminococcaceae (13.8%) and Lachnospiraceae (11.7%), which are consistent previous studies performing 16S rDNA amplicon analyses [40, 78]. Our microbial assembly in the rectum microbiota provided a reference genome resource for future cattle microbiome research using fecal samples.

In cattle fescue toxicosis, the intake of endophyte-infected cool-season grasses produces a significant decrease in animal production due to metabolic and physiological impairments, such as vasoconstriction, reduction in cellulose digestibility, foot root, and other symptoms [38, 84]. For example, Wilbanks *et al.* utilized beef cows in late gestation as the animal model for identifying metabolic perturbance due to fescue toxicity [85], and they reported a reduction in body weight in cows exposed to endophyte-infected tall fescue compared with those exposed to endophyte-free tall fescue [85]. Fescue toxicity leads to lower short-chain fatty acid (SCFA) absorption in the rumen, mainly due to a lower feed intake [86]. Liebe & White (2018) showed in a meta-analysis report that a negative relationship between ergovaline diet concentration and average daily gain [87]. In our toxic tall fescue seed supplement experiments, no treatment × time interaction effect on body weight was detected ($P = 0.13$). However, we did observe a trend for a decreasing body weight due to the addition of K31 seeds during the trial ($P = 0.05$). This tendency in bodyweight reduction is in agreement with previous studies investigating the effect of endophyte-infected tall fescue on animal performance parameters.

Fescue toxicosis alters the gut microbiota through digestive and nutritional changes. Ergovaline could also block monoamine receptors and impairs gut motility, which could perturb the microbes adapted to the gut environment [88, 89]. The influence of fescue toxicosis on cattle gut microbiota was investigated by 16S rDNA ampliconic studies. Mote *et al.* found grazing toxic tall fescue significantly affected the beta-diversity, but not the alpha-diversity of the gut microbiota [17]. Koester *et al.* compared the microbiome of 20 high-tolerate and 20 low-tolerate individuals [40], and discovered a significant reduction in alpha diversity in the low-tolerate group. In this study, we observed a significant reduction in alpha-diversity under fescue toxicosis for all mapped reads (Figure 1B-C) and rumen-microbiota mapped reads (Figure 6D), as well as significant dissimilarity based on beta-diversity before and after treatment (Figure 1D-E), suggesting dysbiosis in the gut microbiota. At the phylum level, Firmicutes were significantly increased, and Proteobacteria and Actinobacteria were significantly reduced after treatment.

Linear discriminant analysis (LDA) effect size (LEfSe) is one of the best approaches for biomarker discovery in the microbiome, which takes into account both the statistical and biological significance [68]. *Ruminococcaceae bacterium P7* is the most significant metagenomic feature that distinguishes the before-treatment and after-treatment microbiome, with an LDA score 5-fold higher than the next best microbial species feature (Figure 3B). *Ruminococcaceae bacterium P7* increased from less than 2% to a quarter of the most abundant family Ruminococcaceae in response to fescue toxicosis, with an over 16-fold enrichment. Ruminococcaceae is also significantly over-represented after treatment as the most significant family feature in the LEfSe analysis. As one of the core rumen microbiota families, Ruminococcaceae is involved in fiber digestion [90], including cellulose and hemicellulose degradation [91]. They are also involved in butyric acid formation in the non-ruminant intestine [92]. 16S metagenomic sequencing has shown that the abundance of many genera within Ruminococcaceae is correlated with average daily gain and production parameters [93]. In this study, WGS metagenomic sequencing provided species and strain level resolution, allowing us to identify that the increase in Ruminococcaceae is driven by *Ruminococcaceae bacterium P7*. There was limited research on this species. Kim *et al.* reported *Ruminococcaceae bacterium P7* as the only enriched taxon at normal environmental conditions compared to heat stress [94] in rumen microbiota, indicating a significant reduction of *Ruminococcaceae bacterium P7* in the rumen under stressed conditions.

Based on our discovery and the results reported in the literature, we hypothesized that the rumen microbiota is impaired after the consumption of endophyte-infected tall fescue seeds, resulting in a global elevation of rumen microbial sequences in the rectum metagenomes (Figure 6A).

Because the increase of rumen core taxa was not uniform (Figure 6B,F), it is not merely a passive process due to the misregulation and loss of the rumen microbial community. Among the enriched microbes after treatment in rectum microbiota, *Ruminococcaceae bacterium P7* is an indicator species with >16-fold changes (Figure 4B), whereas no significant changes were observed in other *Ruminococcaceae bacteria* (Figure 4C). The underlying mechanism of the species-specificity is unclear, which warrants further research. One possible explanation would be that under the fescue toxicosis stress, the rumen *Ruminococcaceae bacterium P7* re-adapted to the environment in the large intestine (i.e., colon), and consequently, a dramatic increase was detected. If this was the case, *Ruminococcaceae bacterium P7* could serve as an excellent biomarker to indicate the degree of fescue toxicosis with the following advantages: first, it could be measured using fecal sampling, which is an easy and affordable technique to obtain representative fecal samples, making large-scale surveys cost-effective; second, the uniformly high fold-change at an individual species level in rectum microbiota will provide great sensitivity and specificity; third, the assay will provide an individual fold enrichment for each animal, which could help determine the proportion of the herd population affected and possibly be used as a selection tool; last but not least, the microbiome changes predate the symptoms, which will inform the management strategies to mitigate fescue toxicosis.

Among the top 41 abundance rumen genera, *Lachnobacterium* increased over 300% after treatment (Figure 6B). This genus has a single know species, *Lachnobacterium bovis*, which is an anaerobic gram-negative species found in cattle rumen and feces [95]. *Lachnobacterium bovis* is significantly enriched in the rectum microbiome after treatment ($P = 0.008$; Figure 3D). The second most increased rumen core genus is *Selenomonas* (Figure 6D), which is a crescent-shaped

Firmicute genus found in the rumen and cecum. At the beginning of the growth phase, most cells are motile with flagella [96]. Sawanon *et al.* found some evidence of its involvement in fiber digestion in rumen [97]. Two species in this genus, *S. ruminantium* and *S. sp. mPRGC8*, were significantly increased after treatment in the rectum microbiome (Figure 3D). Further investigations are needed to determine their role in fescue toxicosis.

Functional pathway enrichment analysis of the rectum microbiome revealed 4/5 most feature KEGG terms before treatment are related to energy metabolism (Figure 7A), which is expected for normal healthy gut conditions. Surprisingly, the overrepresented functional categories shifted to antimicrobial resistance genes and DNA replication proteins (Figure 7A) in response to fescue toxicosis. To the best of our knowledge, changes in the antimicrobial resistome were not reported previously in the literature. Moreover, ARG profiling in the microbiome identified three significantly enriched gene families: tetracycline-resistant ribosomal protection proteins, ABC-F ATP-binding cassette ribosomal protection proteins, and major facilitator superfamily (MFS) antibiotic efflux pumps. The former two are through ribosomal protection mechanisms, and the latter belongs to the largest group of secondary active transporters [98]. In addition, peptidoglycan biosynthesis proteins and bacterial motility proteins (Figure 7B) are also significantly enriched after treatment, which may contribute to the cell wall construction and facilitate the relocation from the rumen to the intestine.

**Availability of data and materials**

N.A.

**Acknowledgments**

# Reference

1. Young CA, Charlton ND, Takach JE, Swoboda GA, Trammell MA, Huhman DV, Hopkins AA: **Characterization of Epichloe coenophiala within the US: are all tall fescue endophytes created equal?** *Front Chem* 2014, **2**.
2. Casler M, Kallenbach R: **Cool season grasses for humid areas**. *Forages The Science of grassland agriculture* 2007, **2**.
3. Leuchtmann A, Bacon CW, Schardl CL, White JF, Tadych M: **Nomenclatural realignment of Neotyphodium species with genus Epichloe**. *Mycologia* 2014, **106**(2):202-215.
4. Hill NS, Pachon JG, Bacon CW: **Acremonium coenophialum-mediated short- and long-term drought acclimation in tall fescue**. *Crop Sci* 1996, **36**(3):665-672.
5. Hill NS, Stringer WC, Rottinghaus GE, Belesky DP, Parrott WA, Pope DD: **Growth, Morphological, and Chemical-Component Responses of Tall Fescue to Acremonium-Coenophialum**. *Crop Sci* 1990, **30**(1):156-161.
6. Clay K: **Fungal Endophytes of Grasses**. *Annu Rev Ecol Syst* 1990, **21**:275-297.
7. Clay K: **The Ecology and Evolution of Endophytes**. *Agr Ecosyst Environ* 1993, **44**(1-4):39-64.
8. Arachevaleta M, Bacon CW, Hoveland CS, Radcliffe DE: **Effect of the Tall Fescue Endophyte on Plant-Response to Environmental-Stress**. *Agron J* 1989, **81**(1):83-90.
9. Read JC, Camp BJ: **The Effect of the Fungal Endophyte Acremonium-Coenophialum in Tall Fescue on Animal Performance, Toxicity, and Stand Maintenance**. *Agron J* 1986, **78**(5):848-850.
10. Bacon CW, Porter JK, Robbins JD, Luttrell ES: **Epichloe-Typhina from Toxic Tall Fescue Grasses**. *Appl Environ Microb* 1977, **34**(5):576-581.
11. Hoveland CS, Haaland RL, King CC, Anthony WB, Clark EM, Mcguire JA, Smith LA, Grimes HW, Holliman JL: **Association of Epichloe-Typhina Fungus and Steer Performance on Tall Fescue Pasture**. *Agron J* 1980, **72**(6):1064-1065.
12. Belesky DP, Stuedemann JA, Plattner RD, Wilkinson SR: **Ergopeptine Alkaloids in Grazed Tall Fescue**. *Agron J* 1988, **80**(2):209-212.
13. Kallenbach RL: **BILL E. KUNKLE INTERDISCIPLINARY BEEF SYMPOSIUM: Coping with tall fescue toxicosis: Solutions and realities**. *J Anim Sci* 2015, **93**(12):5487-5495.
14. Bastin BC, Houser A, Bagley CP, Ely KM, Payton RR, Saxton AM, Schrick FN, Waller JC, Kojima CJ: **A polymorphism in XKR4 is significantly associated with serum prolactin concentrations in beef cows grazing tall fescue**. *Anim Genet* 2014, **45**(3):439-441.
15. Campbell B, Kojima C, Cooper T, Bastin B, Wojakiewicz L, Kallenbach R, Schrick F, Waller J: **A single nucleotide polymorphism in the dopamine receptor D2 gene may be informative for resistance to fescue toxicosis in Angus-based cattle**. *Animal biotechnology* 2014, **25**(1):1-12.
16. Mote RS, Hill NS, Uppal K, Tran VT, Jones DP, Filipov NM: **Metabolomics of fescue toxicosis in grazing beef steers**. *Food Chem Toxicol* 2017, **105**:285-299.
17. Mote RS, Hill NS, Skarlupka JH, Tran VT, Walker DI, Turner ZB, Sanders ZP, Jones DP, Suen G, Filipov NM: **Toxic tall fescue grazing increases susceptibility of the Angus steer fecal microbiota and plasma/urine metabolome to environmental effects**. *Sci Rep* 2020, **10**(1):2497.
18. Sutton JD: **Carbohydrate digestion and glucose supply in the gut of the ruminant**. *Proc Nutr Soc* 1971, **30**(3):243-248.
19. Matthews C, Crispie F, Lewis E, Reid M, O'Toole PW, Cotter PD: **The rumen microbiome: a crucial consideration when optimising milk and meat production and nitrogen utilisation efficiency**. *Gut Microbes* 2019, **10**(2):115-132.
20. Cammack KM, Austin KJ, Lamberson WR, Conant GC, Cunningham HC: **RUMINANT NUTRITION SYMPOSIUM: Tiny but mighty: the role of the rumen microbes in livestock production (vol 96, pg 752, 2018)**. *J Anim Sci* 2018, **96**(10):4481-4481.

21. Ross EM, Moate PJ, Marett LC, Cocks BG, Hayes BJ: **Metagenomic predictions: from microbiome to complex health and environmental phenotypes in humans and cattle**. *PLoS One* 2013, **8**(9):e73056.
22. Shreiner AB, Kao JY, Young VB: **The gut microbiome in health and in disease**. *Curr Opin Gastroenterol* 2015, **31**(1):69-75.
23. McCann JC, Wickersham TA, Loor JJ: **High-throughput Methods Redefine the Rumen Microbiome and Its Relationship with Nutrition and Metabolism**. *Bioinform Biol Insights* 2014, **8**:109-125.
24. Morais S, Mizrahi I: **The Road Not Taken: The Rumen Microbiome, Functional Groups, and Community States**. *Trends Microbiol* 2019, **27**(6):538-549.
25. Zhou M, Peng YJ, Chen Y, Klinger CM, Oba M, Liu JX, Guan LL: **Assessment of microbiome changes after rumen transfaunation: implications on improving feed efficiency in beef cattle**. *Microbiome* 2018, **6**(1):62.
26. Auffret MD, Dewhurst RJ, Duthie CA, Rooke JA, Wallace RJ, Freeman TC, Stewart R, Watson M, Roehe R: **The rumen microbiome as a reservoir of antimicrobial resistance and pathogenicity genes is directly affected by diet in beef cattle**. *Microbiome* 2017, **5**.
27. Wang H, Zheng H, Browne F, Roehe R, Dewhurst RJ, Engel F, Hemmje M, Lu X, Walsh P: **Integrated metagenomic analysis of the rumen microbiome of cattle reveals key biological mechanisms associated with methane traits**. *Methods* 2017, **124**:108-119.
28. Li FY, Guan LL: **Metatranscriptomic Profiling Reveals Linkages between the Active Rumen Microbiome and Feed Efficiency in Beef Cattle**. *Appl Environ Microb* 2017, **83**(9).
29. Li FY, Hitch TCA, Chen YH, Creevey CJ, Guan LL: **Comparative metagenomic and metatranscriptomic analyses reveal the breed effect on the rumen microbiome and its associations with feed efficiency in beef cattle**. *Microbiome* 2019, **7**.
30. Seshadri R, Leahy SC, Attwood GT, Teh KH, Lambie SC, Cookson AL, Eloe-Fadrosh EA, Pavlopoulos GA, Hadjithomas M, Varghese NJ *et al*: **Cultivation and sequencing of rumen microbiome members from the Hungate1000 Collection**. *Nat Biotechnol* 2018, **36**(4):359-367.
31. Delgado B, Bach A, Guasch I, Gonzalez C, Elcoso G, Pryce JE, Gonzalez-Recio O: **Whole rumen metagenome sequencing allows classifying and predicting feed efficiency and intake levels in cattle**. *Sci Rep* 2019, **9**(1):11.
32. Morgavi DP, Kelly WJ, Janssen PH, Attwood GT: **Rumen microbial (meta)genomics and its application to ruminant production**. *Animal* 2013, **7 Suppl 1**:184-201.
33. Huws SA, Creevey CJ, Oyama LB, Mizrahi I, Denman SE, Popova M, Munoz-Tamayo R, Forano E, Waters SM, Hess M *et al*: **Addressing Global Ruminant Agricultural Challenges Through Understanding the Rumen Microbiome: Past, Present, and Future**. *Front Microbiol* 2018, **9**.
34. Hanning I, Diaz-Sanchez S: **The functionality of the gastrointestinal microbiome in non-human animals**. *Microbiome* 2015, **3**.
35. Wallace RJ, Sasson G, Garnsworthy PC, Tapio I, Gregson E, Bani P, Huhtanen P, Bayat AR, Strozzi F, Section FB *et al*: **A heritable subset of the core rumen microbiome dictates dairy cow productivity and emissions**. *Sci Adv* 2019, **5**(7).
36. Li F, Li C, Chen Y, Liu J, Zhang C, Irving B, Fitzsimmons C, Plastow G, Guan LL: **Host genetics influence the rumen microbiota and heritable rumen microbial features associate with feed efficiency in cattle**. *Microbiome* 2019, **7**(1):92.
37. Xue MY, Sun HZ, Wu XH, Liu JX, Guan LL: **Multi-omics reveals that the rumen microbiome and its metabolome together with the host metabolome contribute to individualized dairy cow performance**. *Microbiome* 2020, **8**(1).
38. Melchior EA, Myer PR: **Fescue toxicosis and its influence on the rumen microbiome: mitigation of production losses through clover isoflavones**. *J Appl Anim Res* 2018, **46**(1):1280-1288.

39. Chewning S, Meyer LR, Atchley JA, Powell JG, Tucker JD, Hubbell DS, Zhao J, Kores JE: **Analysis of Fecal Microbiome of Crossbred Beef Cows Grazing Toxic or Novel Fescue.** *J Anim Sci* 2018, **96**:91-91.
40. Koester LR, Poole DH, Serão NVL, Schmitz-Esser S: **Beef cattle that respond differently to fescue toxicosis have distinct gastrointestinal tract microbiota**. *PLoS One* 2020, **15**(7):e0229192.
41. Liu L, Chen X, Skogerbo G, Zhang P, Chen R, He S, Huang DW: **The human microbiome: a hot spot of microbial horizontal gene transfer**. *Genomics* 2012, **100**(5):265-270.
42. Greenblum S, Carr R, Borenstein E: **Extensive strain-level copy-number variation across human gut microbiome species**. *Cell* 2015, **160**(4):583-594.
43. Deurenberg RH, Bathoorn E, Chlebowicz MA, Couto N, Ferdous M, Garcia-Cobos S, Kooistra-Smid AM, Raangs EC, Rosema S, Veloo AC *et al*: **Application of next generation sequencing in clinical microbiology and infection prevention**. *J Biotechnol* 2017, **243**:16-24.
44. Alfaro GF, Rodriguez-Zas SL, Southey BR, Muntifering RB, Rodning SP, Pacheco WJ, Moisa SJ: **Complete Blood Count Analysis on Beef Cattle Exposed to Fescue Toxicity and Rumen-Protected Niacin Supplementation**. *Animals-Basel* 2021, **11**(4).
45. Masiero MM, Roberts CA, Kerley MS, Kallenbach RL: **Evaluation of a commercial genetic test to determine tolerance to fescue toxicity in beef cattle.** *J Anim Sci* 2016, **94**:163-163.
46. Galliou JM, Khanal P, Mayberry K, Poore MH, Poole DH, Serao NVL: **Evaluation of a commercial genetic test for fescue toxicosis in pregnant Angus beef cattle**. *Transl Anim Sci* 2020, **4**(4).
47. Rottinghaus GE, Schultz LM, Ross PF, Hill NS: **An Hplc Method for the Detection of Ergot in Ground and Pelleted Feeds**. *J Vet Diagn Invest* 1993, **5**(2):242-247.
48. Andrews S: **FastQC: a quality control tool for high throughput sequence data**. In*.*: Babraham Bioinformatics, Babraham Institute, Cambridge, United Kingdom; 2010.
49. Bolger AM, Lohse M, Usadel B: **Trimmomatic: a flexible trimmer for Illumina sequence data**. *Bioinformatics* 2014, **30**(15):2114-2120.
50. Li H, Durbin R: **Fast and accurate short read alignment with Burrows-Wheeler transform**. *Bioinformatics* 2009, **25**(14):1754-1760.
51. Sayers EW, Beck J, Bolton EE, Bourexis D, Brister JR, Canese K, Comeau DC, Funk K, Kim S, Klimke W *et al*: **Database resources of the National Center for Biotechnology Information**. *Nucleic Acids Res* 2021, **49**(D1):D10-D17.
52. Li H: **A statistical framework for SNP calling, mutation discovery, association mapping and population genetical parameter estimation from sequencing data**. *Bioinformatics* 2011, **27**(21):2987-2993.
53. Li H, Handsaker B, Wysoker A, Fennell T, Ruan J, Homer N, Marth G, Abecasis G, Durbin R, Genome Project Data Processing S: **The Sequence Alignment/Map format and SAMtools**. *Bioinformatics* 2009, **25**(16):2078-2079.
54. Quinlan AR, Hall IM: **BEDTools: a flexible suite of utilities for comparing genomic features**. *Bioinformatics* 2010, **26**(6):841-842.
55. Comeau AM, Douglas GM, Langille MG: **Microbiome Helper: a Custom and Streamlined Workflow for Microbiome Research**. *mSystems* 2017, **2**(1).
56. Li D, Liu CM, Luo R, Sadakane K, Lam TW: **MEGAHIT: an ultra-fast single-node solution for large and complex metagenomics assembly via succinct de Bruijn graph**. *Bioinformatics* 2015, **31**(10):1674-1676.
57. Pell J, Hintze A, Canino-Koning R, Howe A, Tiedje JM, Brown CT: **Scaling metagenome sequence assembly with probabilistic de Bruijn graphs**. *Proc Natl Acad Sci U S A* 2012, **109**(33):13272-13277.
58. Fu L, Niu B, Zhu Z, Wu S, Li W: **CD-HIT: accelerated for clustering the next-generation sequencing data**. *Bioinformatics* 2012, **28**(23):3150-3152.

59. Menzel P, Ng KL, Krogh A: **Fast and sensitive taxonomic classification for metagenomics with Kaiju**. *Nat Commun* 2016, **7**.
60. Zhu WH, Lomsadze A, Borodovsky M: **Ab initio gene identification in metagenomic sequences**. *Nucleic Acids Res* 2010, **38**(12).
61. Besemer J, Borodovsky M: **Heuristic approach to deriving models for gene finding**. *Nucleic Acids Res* 1999, **27**(19):3911-3920.
62. Oksanen J, Blanchet FG, Kindt R, Legendre P, Minchin P, O'hara R, Simpson G, Solymos P, Stevens M, Wagner H: **Community ecology package**. *R package version* 2013, **2**(0).
63. Shannon CE: **The mathematical theory of communication. 1963**. *MD computing: computers in medical practice* 1997, **14**(4):306-317.
64. Bray JR, Curtis JT: **An Ordination of the Upland Forest Communities of Southern Wisconsin**. *Ecol Monogr* 1957, **27**(4):326-349.
65. Team RC: **R: A language and environment for statistical computing**. 2013.
66. Hadley W: **Ggplot2: Elegant graphics for data analysis**: Springer; 2016.
67. Wickham H: **Reshaping data with the reshape package**. *Journal of statistical software* 2007, **21**(12):1-20.
68. Segata N, Izard J, Waldron L, Gevers D, Miropolsky L, Garrett WS, Huttenhower C: **Metagenomic biomarker discovery and explanation**. *Genome Biol* 2011, **12**(6):R60.
69. Franzosa EA, McIver LJ, Rahnavard G, Thompson LR, Schirmer M, Weingart G, Lipson KS, Knight R, Caporaso JG, Segata N *et al*: **Species-level functional profiling of metagenomes and metatranscriptomes**. *Nat Methods* 2018, **15**(11):962-+.
70. Brunson JC: **Ggalluvial: layered grammar for alluvial plots**. *Journal of Open Source Software* 2020, **5**(49):2017.
71. Ondov BD, Bergman NH, Phillippy AM: **Interactive metagenomic visualization in a Web browser**. *Bmc Bioinformatics* 2011, **12**.
72. Stewart RD, Auffret MD, Warr A, Walker AW, Roehe R, Watson M: **Compendium of 4,941 rumen metagenome-assembled genomes for rumen microbiome biology and enzyme discovery**. *Nat Biotechnol* 2019, **37**(8):953-+.
73. Hunter JD: **Matplotlib: A 2D graphics environment**. *Comput Sci Eng* 2007, **9**(3):90-95.
74. Robinson MD, McCarthy DJ, Smyth GK: **edgeR: a Bioconductor package for differential expression analysis of digital gene expression data**. *Bioinformatics* 2010, **26**(1):139-140.
75. Kanehisa M, Furumichi M, Sato Y, Ishiguro-Watanabe M, Tanabe M: **KEGG: integrating viruses and cellular organisms**. *Nucleic Acids Res* 2021, **49**(D1):D545-D551.
76. Zankari E, Hasman H, Cosentino S, Vestergaard M, Rasmussen S, Lund O, Aarestrup FM, Larsen MV: **Identification of acquired antimicrobial resistance genes**. *J Antimicrob Chemoth* 2012, **67**(11):2640-2644.
77. Alcock BP, Raphenya AR, Lau TTY, Tsang KK, Bouchard M, Edalatmand A, Huynh W, Nguyen ALV, Cheng AA, Liu SH *et al*: **CARD 2020: antibiotic resistome surveillance with the comprehensive antibiotic resistance database**. *Nucleic Acids Res* 2020, **48**(D1):D517-D525.
78. de Oliveira MN, Jewell KA, Freitas FS, Benjamin LA, Totola MR, Borges AC, Moraes CA, Suen G: **Characterizing the microbiota across the gastrointestinal tract of a Brazilian Nelore steer**. *Vet Microbiol* 2013, **164**(3-4):307-314.
79. Li J, Jia H, Cai X, Zhong H, Feng Q, Sunagawa S, Arumugam M, Kultima JR, Prifti E, Nielsen T *et al*: **An integrated catalog of reference genes in the human gut microbiome**. *Nat Biotechnol* 2014, **32**(8):834-841.
80. Xiao L, Feng Q, Liang S, Sonne SB, Xia Z, Qiu X, Li X, Long H, Zhang J, Zhang D *et al*: **A catalog of the mouse gut metagenome**. *Nat Biotechnol* 2015, **33**(10):1103-1108.

81. Pan H, Guo R, Zhu J, Wang Q, Ju Y, Xie Y, Zheng Y, Wang Z, Li T, Liu Z *et al*: **A gene catalogue of the Sprague-Dawley rat gut metagenome**. *Gigascience* 2018, **7**(5).
82. Coelho LP, Kultima JR, Costea PI, Fournier C, Pan Y, Czarnecki-Maulden G, Hayward MR, Forslund SK, Schmidt TSB, Descombes P *et al*: **Similarity of the dog and human gut microbiomes in gene content and response to diet**. *Microbiome* 2018, **6**(1):72.
83. Xiao L, Estelle J, Kiilerich P, Ramayo-Caldas Y, Xia Z, Feng Q, Liang S, Pedersen AO, Kjeldsen NJ, Liu C *et al*: **A reference gene catalogue of the pig gut microbiome**. *Nat Microbiol* 2016, **1**:16161.
84. Schmidt SP, Osborn TG: **Effects of Endophyte-Infected Tall Fescue on Animal Performance**. *Agr Ecosyst Environ* 1993, **44**(1-4):233-262.
85. Wilbanks SA, Justice SM, West T, Klotz JL, Andrae JG, Duckett SK: **Effects of Tall Fescue Endophyte Type and Dopamine Receptor D2 Genotype on Cow-Calf Performance during Late Gestation and Early Lactation**. *Toxins* 2021, **13**(3).
86. Klotz JL: **Activities and Effects of Ergot Alkaloids on Livestock Physiology and Production**. *Toxins (Basel)* 2015, **7**(8):2801-2821.
87. Liebe DM, White RR: **Meta-analysis of endophyte-infected tall fescue effects on cattle growth rates**. *J Anim Sci* 2018, **96**(4):1350-1361.
88. Briejer MR, Mathis C, Schuurkes JA: **5-HT receptor types in the rat ileum longitudinal muscle: focus on 5-HT2 receptors mediating contraction**. *Neurogastroenterol Motil* 1997, **9**(4):231-237.
89. Talley NJ: **Review article: 5-hydroxytryptamine agonists and antagonists in the modulation of gastrointestinal motility and sensation: clinical implications**. *Aliment Pharmacol Ther* 1992, **6**(3):273-289.
90. Deusch S, Camarinha-Silva A, Conrad J, Beifuss U, Rodehutscord M, Seifert J: **A Structural and Functional Elucidation of the Rumen Microbiome Influenced by Various Diets and Microenvironments**. *Front Microbiol* 2017, **8**.
91. La Reau AJ, Suen G: **The Ruminococci: key symbionts of the gut ecosystem**. *J Microbiol* 2018, **56**(3):199-208.
92. Vital M, Karch A, Pieper DH: **Colonic Butyrate-Producing Communities in Humans: an Overview Using Omics Data**. *Msystems* 2017, **2**(6).
93. Mote RS, Hill NS, Skarlupka JH, Turner ZB, Sanders ZP, Jones DP, Suen G, Filipov NM: **Response of Beef Cattle Fecal Microbiota to Grazing on Toxic Tall Fescue**. *Appl Environ Microb* 2019, **85**(15).
94. Kim DH, Kim MH, Kim SB, Son JK, Lee JH, Joo SS, Gu BH, Park T, Park BY, Kim ET: **Differential Dynamics of the Ruminal Microbiome of Jersey Cows in a Heat Stress Environment**. *Animals-Basel* 2020, **10**(7).
95. Whitford MF, Yanke LJ, Forster RJ, Teather RM: **Lachnobacterium bovis gen. nov., sp. nov., a novel bacterium isolated from the rumen and faeces of cattle**. *Int J Syst Evol Microbiol* 2001, **51**(Pt 6):1977-1981.
96. Kingsley VV, Hoeniger JF: **Growth, structure, and classification of Selenomonas**. *Bacteriol Rev* 1973, **37**(4):479-521.
97. Sawanon S, Koike S, Kobayashi Y: **Evidence for the possible involvement of Selenomonas ruminantium in rumen fiber digestion**. *FEMS Microbiol Lett* 2011, **325**(2):170-179.
98. Li XZ, Nikaido H: **Efflux-mediated drug resistance in bacteria: an update**. *Drugs* 2009, **69**(12):1555-1623.
99. Anderson MJ: **A new method for non-parametric multivariate analysis of variance**. *Austral Ecol* 2001, **26**(1):32-46.

## Figure 1. Significant reduction of microbial diversity in cattle rectum microbiota in response to endophyte-infected tall fescue seed supplement.

(**A**) Representative illustration of the experiment design. During the 30-day tall fescue seed supplement, body weight was measured for all eight individuals each week (*orange*: susceptible genotype, N=4; *purple*: tolerant genotype, N=4). Fecal samples from 8 animals were collected at the beginning and the end of this experiment.

(**B-C**) Boxplots of alpha diversity before (*red*) and after (*blue*) treatment observed at the species (**B**) level and genus (**C**) level, measured using the Shannon index. Statistical significance was assessed by a two-sided Wilcoxon signed-rank test.

(**D-E**) The Principal Coordinates Analysis (PCoA) plots of beta diversity between before-treatment and after-treatment rectum microbiota using Bray-Curtis distance (**D**) and Jaccard distance (**E**).

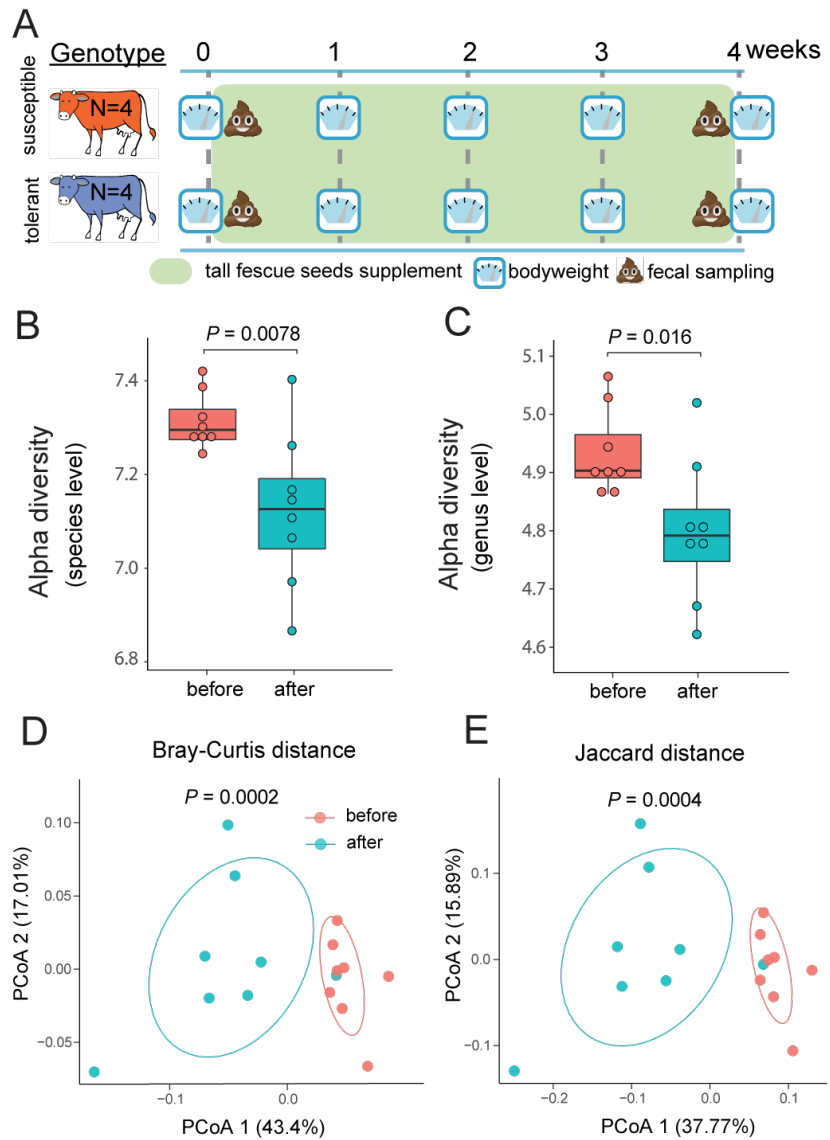

**Figure 2. Assembly and annotation of microbial contigs in cattle rectum microbiota revealed significant taxonomic abundance changes under tall fescue toxicosis.**

(**A**) Relative microbial frequency in the rectum microbiota at the phylum level for the eight animals before (*left bar*) and after (*right bar*) tall fescue seed treatment.

(**B**) Boxplots of frequency for the three most abundant phyla (Firmicutes, Proteobacteria, Actinobacteria) showing significant changes in relative frequency after tall fescue seed treatment (two-sided Wilcoxon signed-rank test).

(**C-D**) Heatmap of relative frequency for the top 20 most abundant bacteria families (**C**) and species (**D**). The taxa names were rank-ordered with the most abundant taxon on the left.

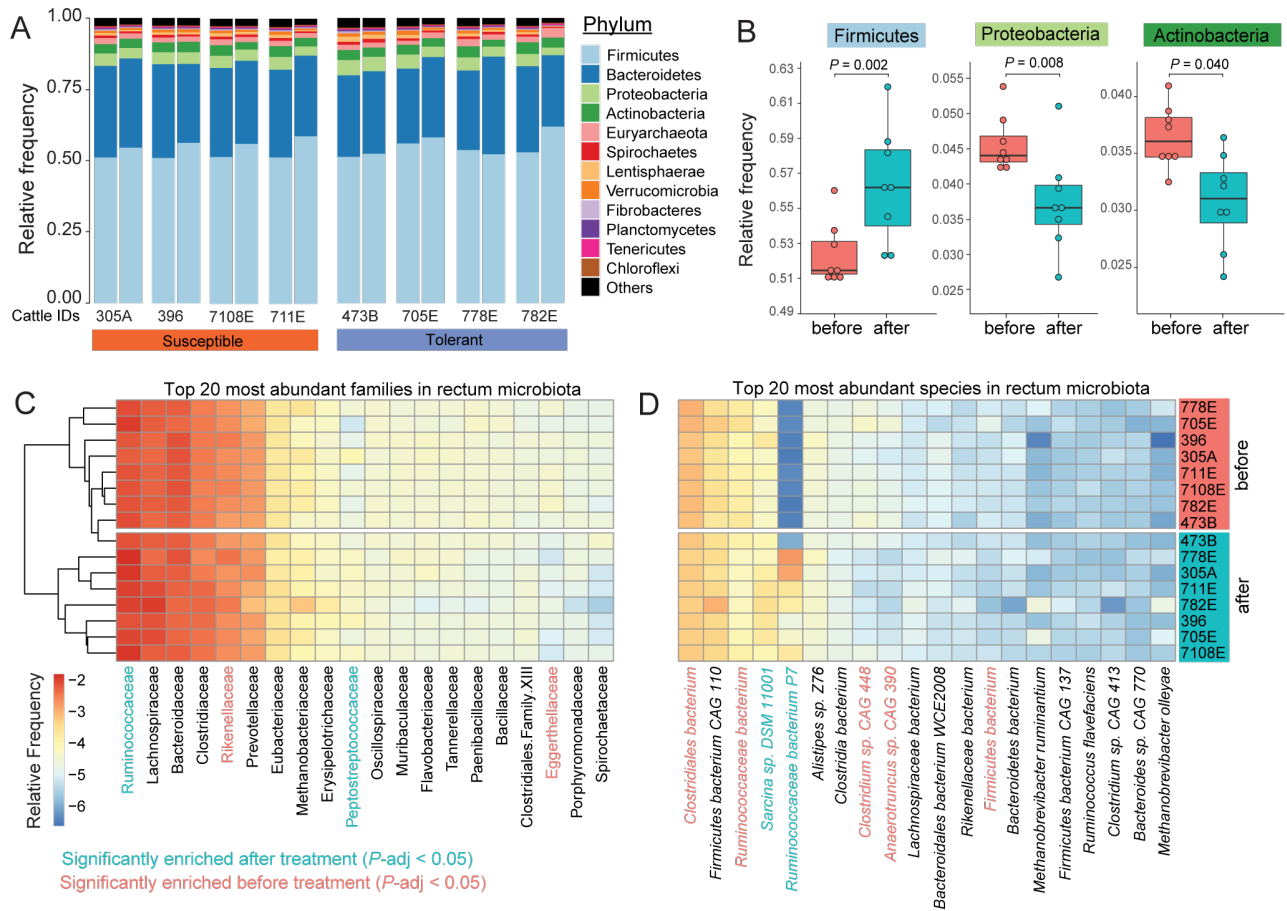

**Figure 3. Significant changes in taxonomic abundance at family and species level after tall fescue seed consumption.**

(***A-B***) The linear discriminant analysis (LDA) score of top 10 most featured families (***A***) and species (***B***) in cattle rectum microbiota before and after tall fescue seed treatment (***red***: enriched before treatment; ***blue***: after treatment).

(***C***) Alluvial plot of the top 10 featured species in response to tall fescue seed treatment (***orange***: susceptible genotype; ***purple***: tolerant genotype).

(***D-E***) Heatmaps of the relative frequency for the significantly increased (***D***) and decreased (***E***) species in cattle rectum microbiota after treatment. The species names are shown on the right.

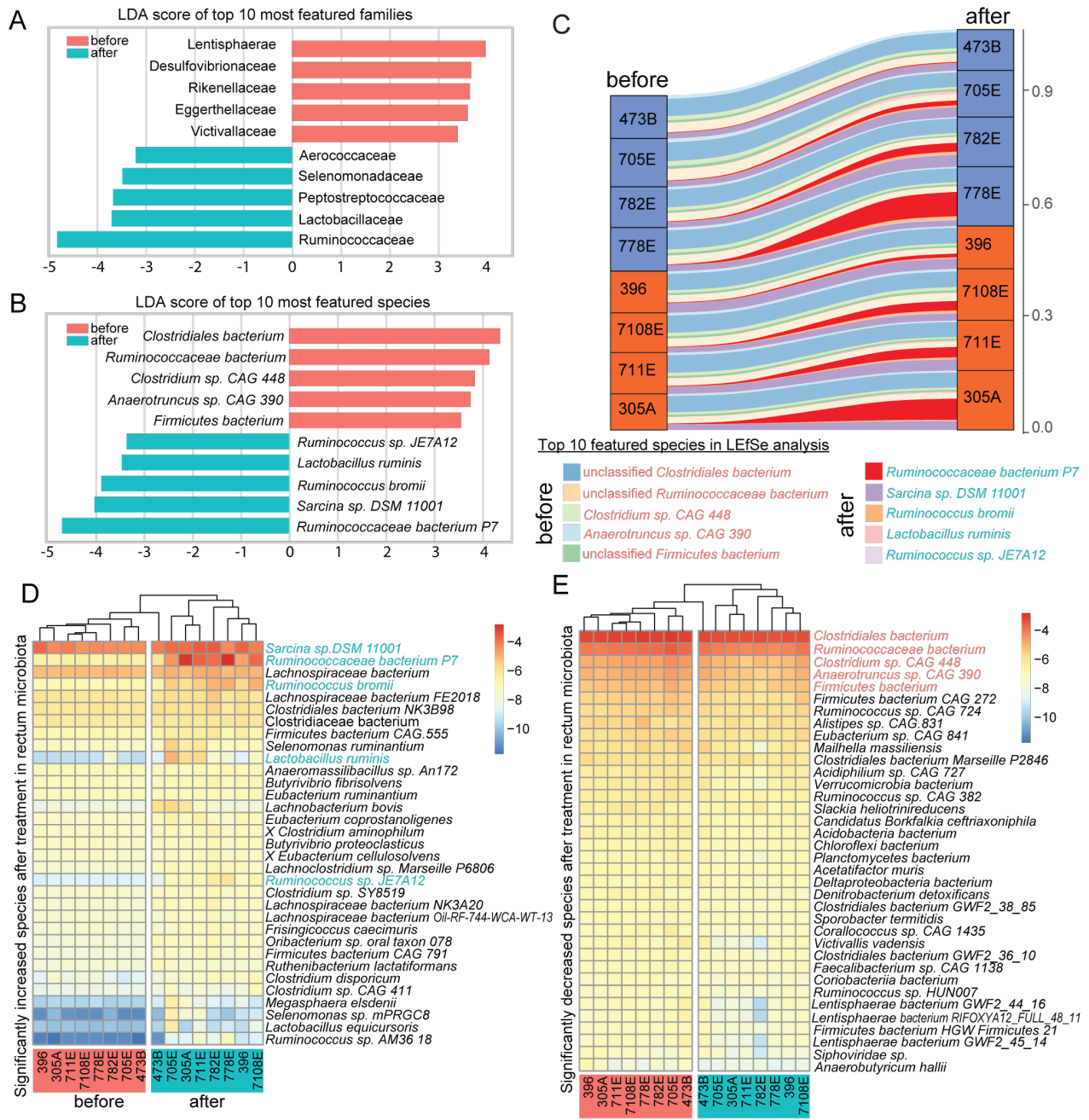

**Figure 4. The abundance of a specific Ruminococcaceae species, *Ruminococcaceae bacterium P7*, increased over tenfold in cattle rectum microbiota after treatment.**

(**A**) Barplot of the species-level relative frequency in cattle rectum microbiota before (*left bar*) and after (*right bar*) tall fescue seed treatment. The species names and color legends were shown for the top 20 most abundant species.

(**B**) A boxplot of the relative frequency before (*red*) and after (*blue*) treatment for Ruminococcaceae bacterium P7, showing a significant increase in abundance (two-sided Wilcoxon signed-rank test).

(**C**) Heatmap of the relative frequency for 17 Ruminococcaceae bacterium species in cattle rectum microbiota.

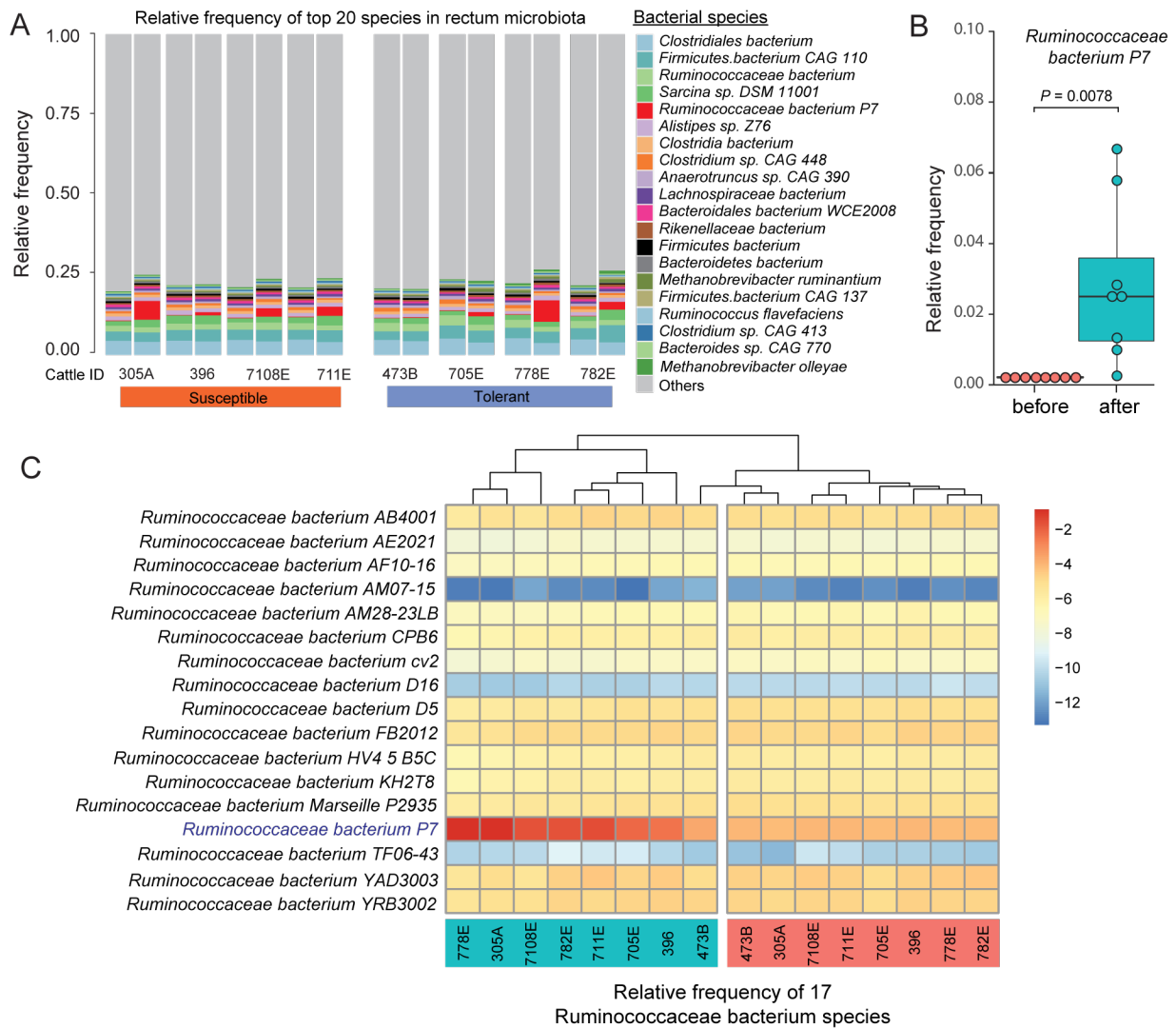

**Figure 5. Krona phylogenetic piechart illustrating the composition and phylogenetic relationship of taxonomy units nested within Firmicutes and Ruminococcaceae.**

(**A-B**) Annotated taxonomy units within Firmicutes phylum were visualized in terms of relative abundance and taxonomic hierarchy for cattle rectum microbiota before (**A**) and after (**B**) treatment. Different taxonomic terms are color-coded, and the composition percentages are labeled at the family level. The area in the chart is proportional to the relative frequency.

(**C-D**) Annotated taxonomy units within the Ruminococcaceae family were visualized in terms of relative abundance and taxonomic hierarchy for cattle rectum microbiota before (**C**) and after (**D**) treatment. Different taxonomic terms are color-coded, and the composition percentages are labeled at the species level. The area in the chart is proportional to the relative frequency.

The relative proportions of *Ruminococcaceae bacterium P7* are labeled in all four panels.

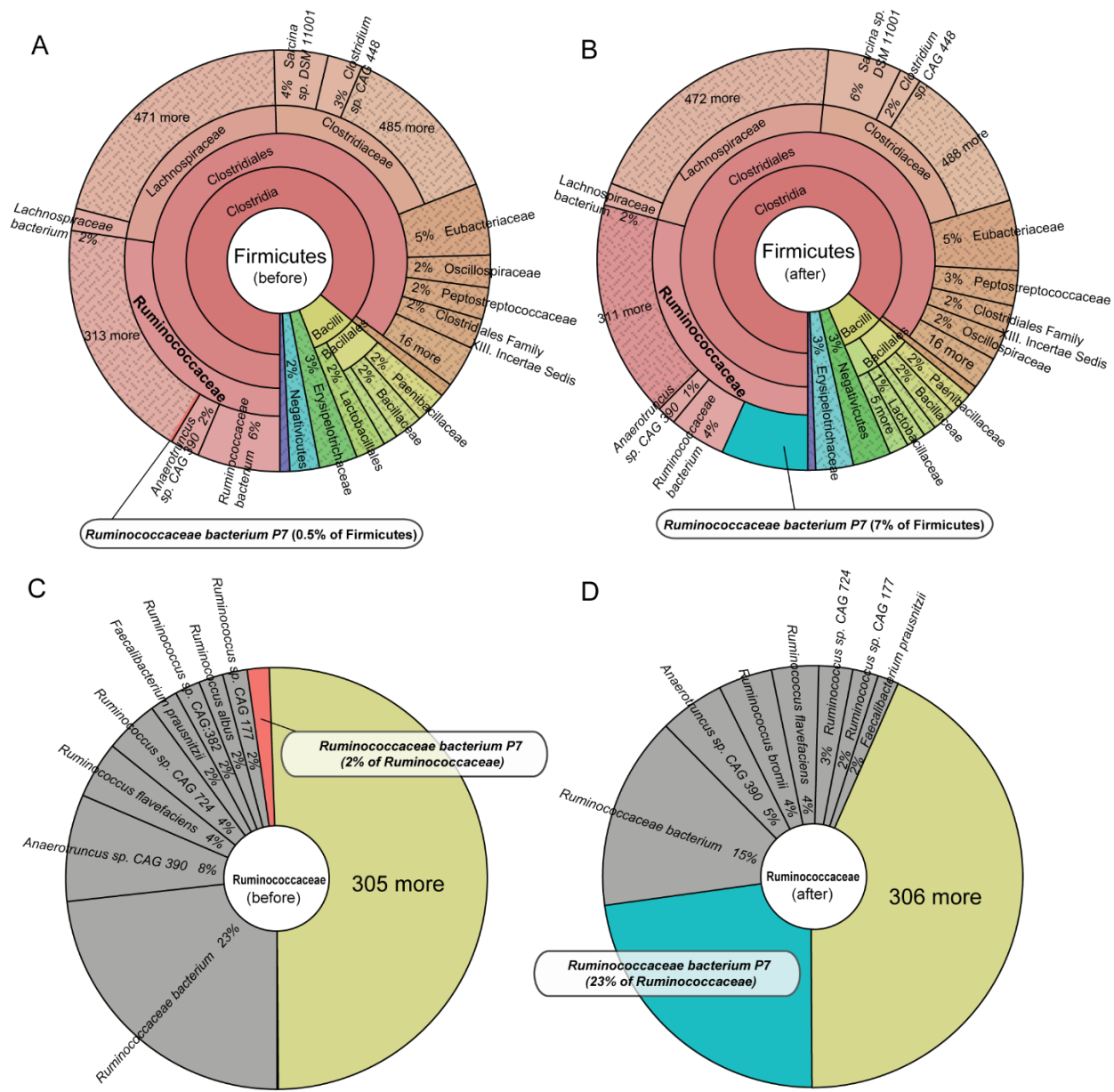

**Figure 6. Comparative metagenomic analysis of cattle rectum microbiota before and after treatment with the rumen reference microbiome profile.**

(**A**) A boxplot of the alignment percentage of rectum reads to the rumen reference metagenomic assembly [72] ($P$-value = 0.078; two-sided Wilcoxon signed-rank test).

(**B**) A barplot of the relative change of 41 core rumen microbiota genera [72] in rectum microbiota after tall fescue treatment (***blue***: increase, ***red***: decrease).

(**C**) Venn diagram of the overlap between top 41 rumen genera and top 41 rectum core genera (***orange***: rumen core genus, ***blue***: rectum core genus).

(**D**) Boxplots of alpha diversity measured by the Shannon index ($P$-value = 0.04, two-sided Wilcoxon signed-rank test), using the rectum abundance of the top 31 rumen families before and after treatment.

(**E**) PCoA plot of beta diversity at species abundance level before and after treatment using Bray-Curtis distance ($P$-value = 0.0004, PERMANOVA [99]).

(**F**) Illustration of the top 31 most abundant rumen microbiota families and their abundance in rectum microbiota. ***Top:*** the abundance of the top 31 rumen microbiota family calculated using MetaBAT2 in rumen-uncultured genomes in [72] plotted in log10 scale. ***Bottom:*** heatmap of the relative frequencies of corresponding families in the 16 cattle rectum metagenomes before and after treatment. The families are ordered based on average rectum microbiota abundance from high to low.

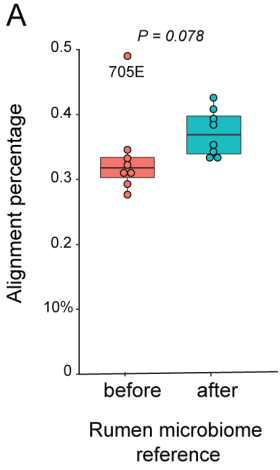
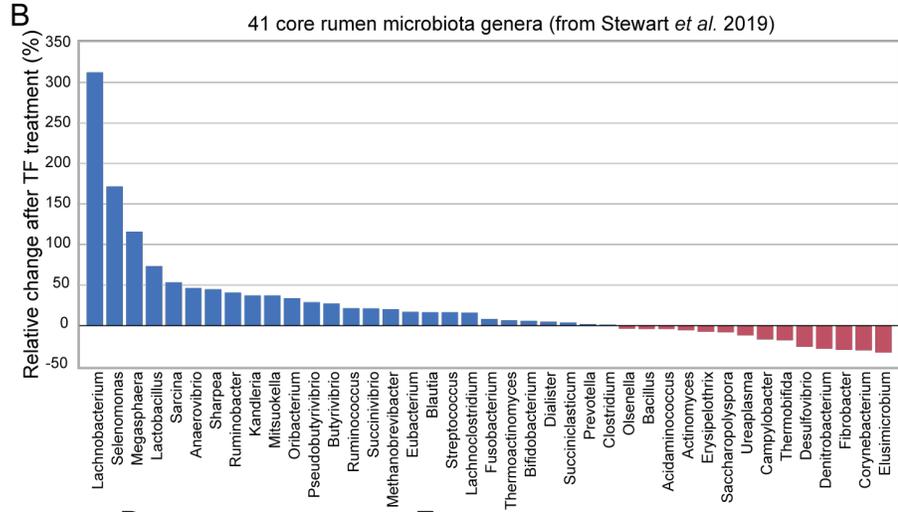
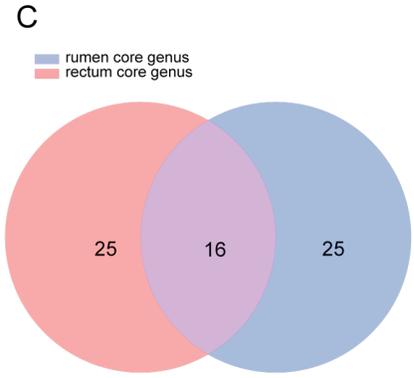
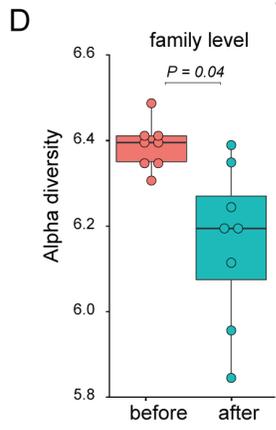
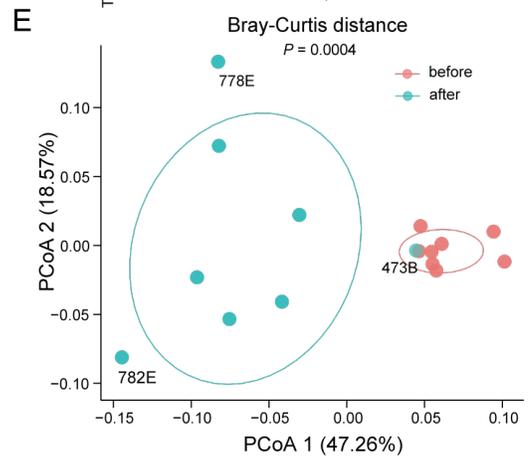
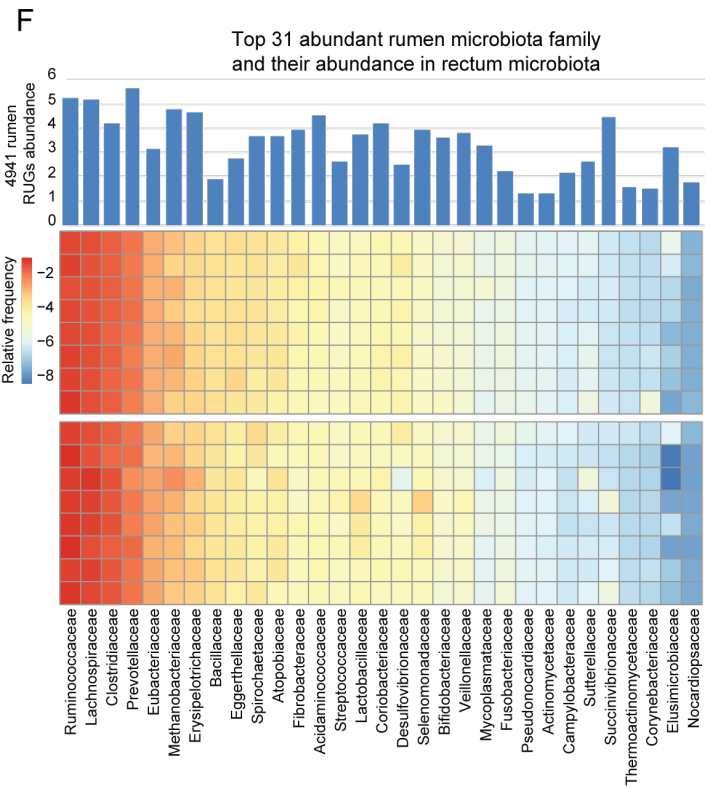

**Figure 7. Tall fescue toxicosis resulted in an increase in both the number and relative abundance of ARGs in the antimicrobial resistome.**

(**A**) LDA score of top 10 most featured KEGG terms in cattle rectum microbiota before and after TF treatment (*red*: before; *blue*: after).

(**B**) The barplot of normalized relative abundance (counts-per-million-reads, CPM) before (red) and after treatment (blue) of all KEGG BRITE hierarchy (*, $P < 0.05$; **, $P < 0.01$; two-sided Wilcoxon signed-rank test). KEGG BRITE terms are labeled according to their functional category: metabolism (gold), genetic information processing (green), and signaling and cellular processes (black).

(**C**) Heatmap of the CPM of identified antimicrobial resistance gene families with column cluster.

(**D**) Boxplots of 3 significantly increased antimicrobial resistance gene families (major facilitator superfamily antibiotic efflux pump; tetracycline resistant ribosomal protection protein; ABC-F ATP-binding cassette ribosomal protection protein; two-sided Wilcoxon signed-rank test).

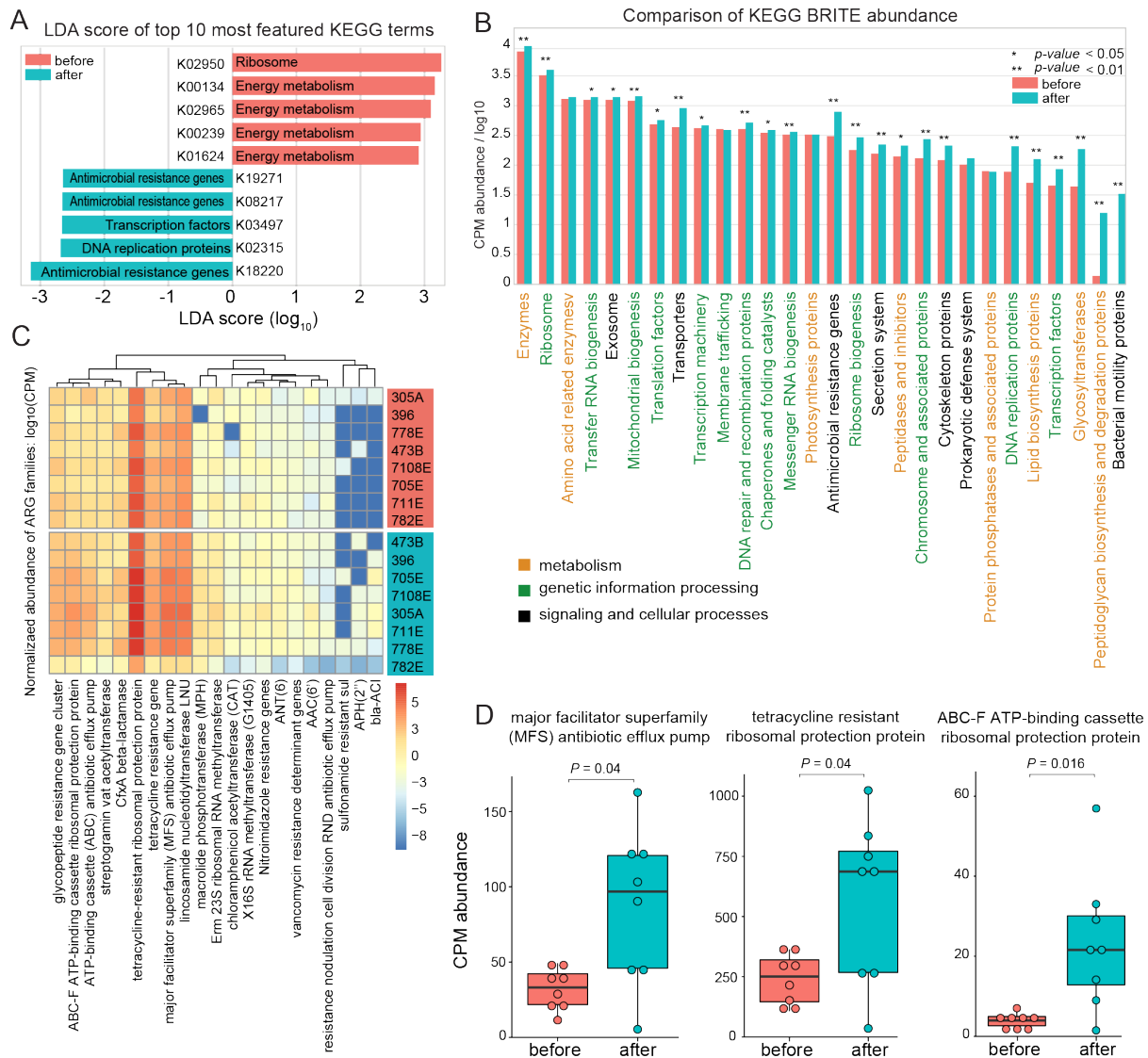